\documentclass[aip,jcp]{revtex4-1}

\usepackage{graphicx}
\usepackage{dcolumn}
\usepackage{bm}

\usepackage[dvips]{epsfig}
\usepackage{float}

\usepackage{docs}
\usepackage[sort&compress]{natbib}

\begin{document}
\baselineskip = 24pt

\title{Spherical Polymer Brushes Under Good Solvent Conditions: Molecular Dynamics Results Compared to Density Functional Theory} 



\author{Federica Lo Verso}
\email[]{loverso@uni-mainz.de}
\affiliation{Institut f{\"u}r Physik, Johannes-Gutenberg-Universit\"at Mainz, D-55099 Mainz, Germany.}
\author{Sergei A. Egorov}
\affiliation{Department of Chemistry, University of Virginia, Charlottesville VA22901, USA}
\author{Andrey Milchev}
\affiliation{Institute for Physical Chemistry, Bulgarian Academy of Sciences, Sofia, Bulgaria}
\author{Kurt Binder}
\affiliation{Institut f{\"u}r Physik, Johannes-Gutenberg-Universit\"at Mainz, D-55099 Mainz, Germany.}
%

\date{March 2010}
\revised{\today}%

\begin{abstract}
A coarse grained model for flexible polymers 
end-grafted to repulsive spherical nanoparticles is studied for various chain lengths and grafting densities under good solvent conditions, by Molecular Dynamics methods and density functional theory.
With increasing chain length the monomer density profile exhibits a crossover to the star polymer limit.
The distribution of polymer
 ends and the linear dimensions of individual polymer chains are obtained,
while the inhomogeneous stretching of the chains 
is characterized 
by the local persistence lengths.
The results on the structure factor of both single 
chain and full spherical brush 
as well as the range of applicability of the different theoretical tools are presented.
Eventually an outlook on experiments is given.

\end{abstract}


\maketitle


\section{Introduction}

\label{intro:sec}

Polymer chains 
densely grafted by a special endgroup to a substrate surface on which they otherwise do not adsorb, 
stick away from the substrate, forming a polymer brush. 
\cite{Milner_1991,Halperin_1991,Grest_1995,Szleifer_1996,Grest_1999,Leger_1999,PolymerBrushes_2004,Dimitrov_2007}
Polymer brushes play a role of great scientific and  technological relevance, even beyond colloidal stabilization.\cite{PolymericStabilization_1983,PolymerAdsorption_1984}
The applications range from lubrication, tuning of adhesion and wetting properties, to biotechnology including protein purification, enzyme immobilization, virus capturing, improvement of biocompatibility of drugs, etc.\cite{Klein_1996,Klein_2009,PolymerBrushes_2004,Brown_1996,*Brown_1993,Lee_2006,Storm_1995}
The rich variety of practical uses in our daily life depends on the specific architecture and intramolecular interactions
via  several parameters, such as molecular weight of the chains, surface density, matching between 
the properties of the solvent and the monomeric units of the polymer.
Due to the interplay 
between 
excluded volume interactions among the chains 
and the entropic repulsion of 
the grafting substrate, polymer brushes acquire 
nontrivial structures where a polymer conformation is characterized by multiple length scales.\cite{Binder_2002}
In this context 
the study of
flat planar grafting surfaces has 
found 
longstanding interest,
from the point of view of statistical mechanics of the macromolecular configurations (see e.g. Refs.\citenum{Alexander_1987,deGennes_1980,Skvortsov_1988,Cosgrove_1987,Milner_1988,Murat_1989,Chakrabarti_1990,Muthukumar_1989,Zhulina_1990,Lai_1991,Wittmer_1994,Netz_1998,Kreer_2004}).
For many applications 
(building blocks of nanocomposites \cite{Vaia_2007,Hall_2010,Dukes_2010}, surface modification of biomolecules 
\cite{Storm_1995,Trombly_2009} etc.) the grafting surfaces are (approximately) spherical, with a radius comparable to the linear dimensions of the grafted polymers.
However the effect of a spherical geometry of the
substrate on the brush structure
has been considered mostly for the extremely high curvature regime.
It is indeed well known that
 when the radius of the central nanoparticle gets smaller than the typical size of the free macromolecule
in solution, a crossover to the behavior of star polymers\cite{Roovers_1972,Daoud_1982,Burchard_1983,Birshtein_1984,Birshtein_1986,Duplantier_1986,*Duplantier_1986_b,Grest_1987,Grest_1994} is expected.
The intermediate curvature regime, where the size of the core is comparable to the full molecule extension,
has received somewhat less attention.\cite{Dukes_2010,Ball_1991,Zhulina_1991,Dan_1992,Wijmans_1993,Toral_1993,McConnell_1994,Lindberg_2001}
Another debated 
aspect is related to the 
description of 
micelles \cite{Foenster_1996} in terms of 
spherical polymer brushes.
Small spherical nanoparticles with many flexible chains grafted to their surface are similar to spherical micelles, formed from asymmetric block copolymers 
$A_{l}B_{1-l}$ with $l<<$1 in a selective solvent (which is bad for the A-block).
The A-monomers form a dense core, and the A-B interface can be idealized as a sphere, from which the B-blocks stick out \cite{Hong_1980,Leibler_1983,Leibler_1984,Zhulina_1985,Halperin_1987,Munch_1988,Nagarajian_1989,Bhattacharya_2001}.
In these models both the shape fluctuations of the A-rich core and the nonzero width of the A-B surface are ignored.
Thus, an improved understanding of spherical brushes may be useful both for various applications of these nanoparticles, and to 
elucidate 
properties of related systems such as spherical micelles.

In the present work, we
study an idealized coarse-grained model for spherical polymer brushes both by Molecular Dynamics (MD) computer simulations and by calculations using density functional theory.\cite{Egorov_2005,*Patel_2005,*Egorov_2007,*Egorov_2007_b,*Striolo_2007}
We have $f$ flexible polymer chains containing $N$ effective beads, described by the standard Kremer-Grest \cite{Grest_1986,Kremer_1990} model.
These chains are grafted to sites regularly distributed on the surface of spheres such that the values of $\sigma$=0.068, 0.118 and 0.185
of the grafting density are realized.
The motivation for choosing these values originates from the fact that the same model and parameters have 
already been studied for planar brushes.\cite{Dimitrov_2006} 
The main purpose of the present work is to give a detailed quantitative description of the spherical brush behavior when the radius of gyration is comparable with the size of the core.
At the same time, by changing the length of the polymer chains
we explore the crossover between the flat brush regime (small 
curvature, short chains) and the star polymer regime (big 
curvature, long chains).
More details on the model and simulation technique will be given in the following section (\ref{section2}).
In Sec.\ref{section3}, we describe our numerical results, while Sec.\ref{section4} presents a discussion where comparison of MD to density functional calculations as well as simulation results for flat planar brushes are given. 
Eventually,
referring to 
simulations and experimental literature
we 
discuss the issue of which features of our results are universal or 
model-dependent.
Sec.\ref{section5} presents our conclusions.

\section{Some Comment on the Model and Simulation Techniques}
\label{section2}

In order to realize the grafting densities $\sigma$=0.068, 0.118, 0.185 \cite{Dimitrov_2006} on spheres with a hard-core radius $R_c$ that is of comparable size to the gyration radius $R_{g}$ of the chain,
we decided to choose respectively $f$=42 and $R_{c}$=7, $f$=92 and $R_{c}$=7.9, $f$=162 and $R_{c}$=8.35.
With this choice of parameters $R_{g}$ can be conveniently calculated in our simulation.
Here all the lengths are 
measured in units of the Lennard-Jones diameter ${\sigma}_{LJ}$=1 of the beads that form the chains.

For planar brushes it is known \cite{Lai_1991} that it makes little difference whether the grafting sites are regularly arranged at the grafting surface or at random, according to the chosen value of $\sigma$.
However to average out sample to sample fluctuations, the random distribution requires averaging over a large number of realization of grafting sites.
To avoid unnecessary huge computation efforts, 
we decided to consider a regular distribution of 
the grafted monomers only.
The method to find the appropriate positions of these grafting sites is a geodesic subdivision, i.e. a repeated subdivision of triangles.
According to this approach one may begin with an icosahedron inscribed in a unit sphere, find the midpoint of each edge and renormalize its coordinates, pushing it out to obtain a new vertex lying on the unit sphere.
This divides each original triangle into four new smaller triangles.
It is possible to repeat this process and thus increase the number of vertices.
One should note, however, that the lengths of the edges connecting the vertices are not all exactly the same, because of the projection to the unitary sphere.
When the desired 
number of grafting sites is reached, the corresponding value of $R_c$ is readily 
calculated from 
$\sigma$.

The first monomer of each chains is rigidly fixed to the core.
All the monomers interact
via  a truncated and shifted Lennard-Jones (LJ) potential:

\begin{equation}\label{LJ}
V_{\mathrm{LJ}}(r)=\left\{
\begin{array}{cccc}
 & 4 {\varepsilon}_{\mathrm{LJ}}\left[\left(\frac{{\sigma}_{\mathrm{LJ}}}{r}\right)^{12}-\left(\frac{\sigma_{{\mathrm{LJ}}}}{{r}}\right)^6 +\frac{1}{4}\right] & \mathrm{for} & r\leq 2^{1/6} \sigma_{\mathrm{LJ}}\\
 & 0 &  \mathrm{for} & r> 2^{1/6} {\sigma}_{\mathrm{LJ}}
\end{array}
\right.
\end{equation}
where $r$ is the distance between the beads, 
while $\sigma_{\mathrm{LJ}}$(=1) and $\varepsilon_{\mathrm{LJ}}$(=1) set the scales
for length and energy respectively.
The chains connectivity then is modelled by the 
finitely extendable nonlinear elastic (FENE) potential.\cite{Grest_1987,Grest_1994,Grest_1986,Kremer_1990}
%
\begin{equation}\label{FENE}
V_{\mathrm{FENE}}(r)=\left\{
\begin{array}{cccc}
 & \frac{k_{{\mathrm{FENE}}}}{2}\left(\frac{R_0}{\sigma_{\mathrm{LJ}}}\right)^{2} ln\left[1-\left(\frac{r}{{R_0}}\right)^2 \right] & \mathrm{for} & r \leq R_0\\
 & \infty & \mathrm{for} &  r>R_0
\end{array}
\right.
\end{equation}
Here $k_{\mathrm{FENE}}=30\varepsilon_{\mathrm{LJ}}$ and
$R_0=1.5{\sigma}_{LJ}$ as usual.\cite{Grest_1987,Grest_1994,Grest_1986,Kremer_1990}
For this choice the total potential between bonded monomers ($V$=$V_o$$+V_{{\mathrm{FENE}}}$) 
has a minimum at $r$$\sim$$0.97\sigma_{\mathrm{LJ}}$.
Eq.\ref{LJ} mimics good solvent conditions (the solvent molecules are not taken explicitly into account).
In our analysis we considered several chain lengths:
 $N$=20,40,60 and 80 (the values do not take into account the grafted monomers).

\begin{figure}
\includegraphics[width=9.0cm,clip]{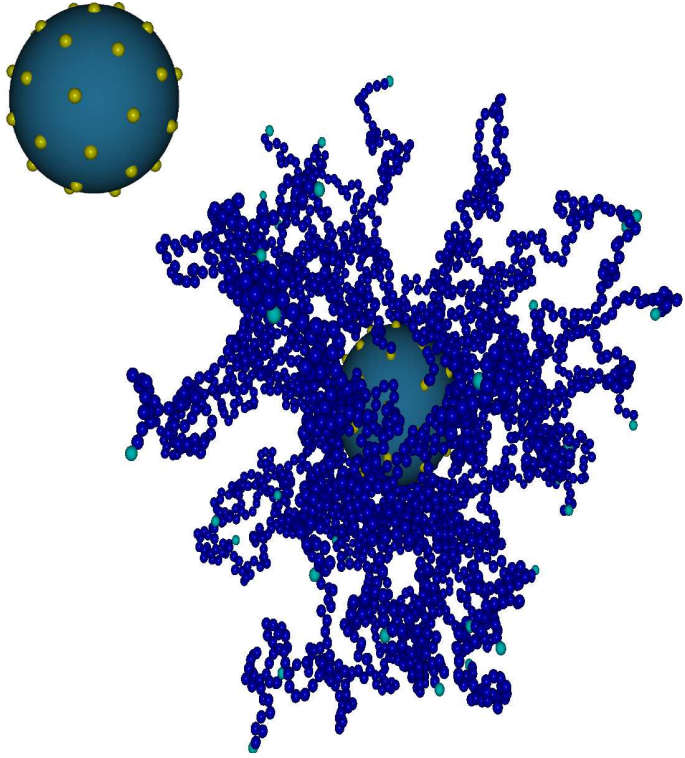}
\caption{\label{fig-1}Snapshot of a spherical brush with $f$=42, $\sigma$=0.068 and $N$=$80$.
The additional grafted monomer is shown in light gray (yellow online). 
On the top (left side): 
the distribution of the grafted monomers on the sphere is shown. 
The free end monomers are highlighted as grey spheres (light blue online).}
\end{figure}

Standard MD methods using the velocity Verlet algorithm \cite{Allen_1987} were used, following previous works where multi-arm star polymers were simulated.\cite{Grest_1987,Grest_1994,Jusufi_1999}
In order to stabilize
the temperature of the system a Langevin thermostat as described in Refs. \citenum{Grest_1987,Grest_1994,Grest_1986,Kremer_1990} is implemented
and the chosen temperature is $k_{B}T=1.2$.
Assigning to the monomer mass $m$ the value $m=1$, the characteristic time $\tau = \sqrt{m\sigma_{\rm \mathrm{LJ}}^2/\varepsilon_{\mathrm{LJ}}}$
is then unity too.
The integration time step
was covering the range $[10^{-3}\tau,10^{-2}\tau]$ with a total number of $1$-$2$$\times$$10^6$ timesteps
used for equilibration and several 
independent runs 
of $3$-$5$$\times$$10^7$ timesteps each to gather statistics.
Fig.\ref{fig-1} represents an 
illustrative example for the configurations thus generated.
In the picture we shows a snapshot 
of a spherical polymer brush with $f=42$,
$\sigma$=0.068, and N=80.

\section{Simulation Results}
\label{section3}

\begin{figure}
\includegraphics[width=12.0cm,clip]{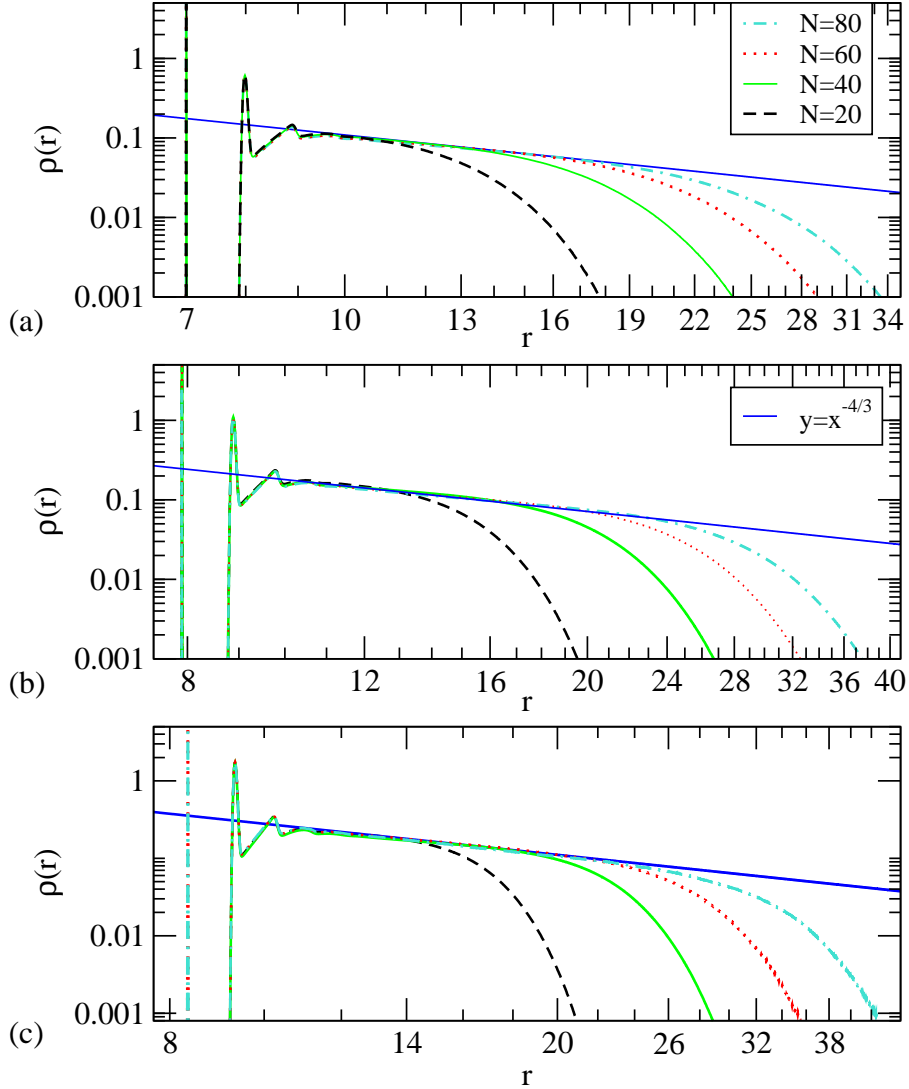}
\caption{\label{fig2}Log-log plot  for the monomer density $\rho(r)$ at distance $r$ from the center of the sphere
of a spherical brush with $f$=42 (a), 92 (b) and 162 (c).
The corresponding sphere radii and grafting densities are
$R_c$=7.0, $\sigma$=0.068 (a), $R_c$=7.9, $\sigma$=0.118 (b) and $R_c$=8.35, $\sigma$=0.185 (c) respectively. Four choices of chain length, $N$=20, 40, 60,80,
are plotted in each figure, as indicated.
The straight line illustrates the Daoud-Cotton \cite{Daoud_1982} power law proposed for star polymers $\rho(r)\propto r^{-4./3.}$.}
\end{figure}

We start our study analysing the conformational properties of the molecule.
Fig.\ref{fig2} shows the log-log plots for the radial monomer density profiles. Note that the grafted monomer at $r$=$R_c$ shows up as a delta-function spike.
%
The pronounced structure close to the core is evidenced even further.
The first mobile monomer (at about $r$=$R_c$+$r_{bond}$, $r_{bond}$ $\sim$0.97) is tightly bound to the grafting surface and hence shows up as a rather sharp peak.
A 
further peak for the second monomer is also clearly seen, while thereafter (for $N$$\geq$ 40) already the characteristic power law decay ${\rho(r)\propto r^{-4/3}}$ expected for star polymers with very large arms \cite{Daoud_1982} sets in.
Apart from a shift of the abscissa scale
the data closely resemble simulation results for star polymers \cite{Grest_1987,Grest_1994}.

The pronounced peaks near 
$r$=$R_c+r_{bond}$, are reminiscent of the layering of particles in fluids near hard walls, and this feature of our results is similar to the data for the density of polymer brushes grafted to flat planar walls.\cite{Murat_1989,Kreer_2004,Dimitrov_2006}
Of course in the region where the layering has died out, for flat polymer brushes one observes a decay of 
$\rho(r)$ roughly compatible with a parabolic decay \cite{Milner_1988} rather than the Daoud-Cotton power law 
observed here.
For low enough $N$ and/or big enough radius of the internal core, 
one should expect 
a crossover to the flat brush limit.
As we will show in detail in Sec.\ref{section4}
the results we obtained for $N$$<$40  evidenced a 
parabolic profile.

Finally, when the
radial distance is $r\sim R_c+R_{gr}(N)$ we observe a crossover of $\rho(r)$ to an exponential decay ($R_{gr}(N)$ here is the gyration radius component of a chain in the radial direction.)
\begin{figure}
\includegraphics[height=13.0cm,clip]{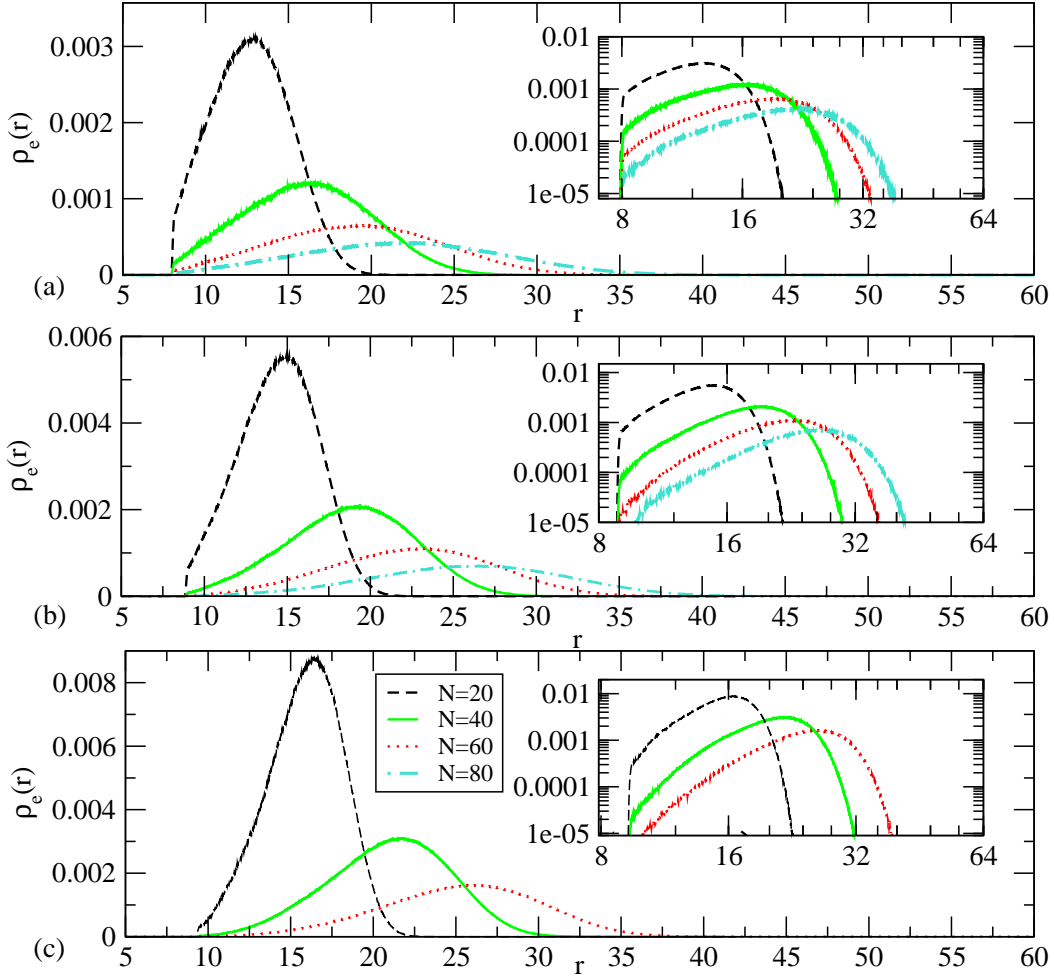}
\caption{\label{fig3}Linear-linear plots of the free end monomer density profiles $\rho_e(r)$, versus the radial distance $r$ for the case $f$=42(a), 92(b), 162(c) and several chain lengths, as indicated.
Insets (a), (b), (c) show the same data on log-log plots.}
\end{figure}

Fig.\ref{fig3} shows the corresponding density profile of the free chain ends.
Unlike the Daoud-Cotton \cite{Daoud_1982} picture, chain ends do not only occur in the outer region of the spherical polymer brush, but rather 
throughout the whole brush.
We point out here that for 
star polymers, scaling theories and renormalization group predictions \cite{Duplantier_1986,*Duplantier_1986_b,Ohno_1991} imply
\begin{equation}\label{densprof}
\rho_{e}(r)\propto r^{\theta(f)},  r_{bond}<<r<<R_{gr}(N),
\end{equation}
%
The exponent $\theta$ can be expressed in terms of the exponents $\gamma$ and $\nu$ of linear polymers \cite{deGennes_1979}
and the partition function exponent $\gamma(f)$ of $f$ arm star polymers as \cite{Ohno_1991}
\begin{equation}\label{theta} 
\theta(f)=[\gamma -\gamma(f+1)+\gamma(f)-1]/\nu.
\end{equation}
%
Since for large $f$ one expects $\gamma(f)\propto-f^{3/2}$, Eq. \ref{theta}
would imply a very large exponent, $\theta(f)\propto f^{1/2}/\nu$, which practically means that the chain ends are excluded from a substantial fraction of the internal corona
of 
the star polymer. In contrast, for a brush on a planar grafting surface one finds from the self-consistent field theory in the strong stretching limit \cite{Milner_1988}
\begin{equation}\label{densprof_plan}
\rho_{e}(z)\propto z,   r_{bond}<<z<<R_{gz}(N),
\end{equation}
where $z$ is the coordinate perpendicular to the grafting surface. Unfortunately, $R_{gr}(N)$ for the available chain lengths is clearly 
not large enough to provide a significant regime of $r$ where Eq.\ref{densprof} could be tested. The pronounced curvature of $\rho_e(r)$ on the log-log plots (inset in figure \ref{fig3}) prevents us from making any strong statement on the exponent $\theta (f)$.
In fact, on general grounds, one would expect a crossover from a linear behavior, when $f$ is very large and $N$ relatively short (such that $R_{gr}(N)<<R_{c}$) and the system locally resembles a polymer brush on a flat surface, to a very steep power law (Eq.\ref{densprof}) in the inverse limit, where $R_{gr}(N)>>R_{c}$ and the system resembles a star polymer.
Our data, of course, can probe only a small intermediate part of this very extended crossover regime. 
Qualitatively, our results for $\rho_{e}(r)$ are similar to the results by Toral and Chakrabarti\cite{Toral_1993} for a different model (the so-called pearl-necklace model).
We 
recall that for star polymers in $d=2$ dimensions where $\theta(f)$ is known exactly, since \cite{Duplantier_1986,*Duplantier_1986_b}
$\gamma (f)$=17/16+(9/32)$[f-f(f-1)/2]$ is known from conformal invariance methods, Monte Carlo results for star polymers even for small $f$ did 
not confirm Eqs.~\ref{densprof},\ref{theta}.\cite{Ohno_1991}
Clearly, it would be desirable to vary both $f$ and $N$ over a much wider range, but this is still impossible with the present method.

Another issue of interest is the local stretching of the chains.
Since the end-to-end vector $\overrightarrow{R_e}$ of the chains, by the symmetry of the problem, is oriented on average in the perpendicular direction, we monitored the local persistence length $l_p(k)$ of the chains, 
defined as \cite{Flory_1969}
\begin{equation}\label{pl}
l_{p}(k)/r_{bond}=\langle \overrightarrow{a_k}\cdot\overrightarrow{R_e}/|\overrightarrow{a_k}|^2\rangle, \overrightarrow{a_k}=\overrightarrow{r_k}-\overrightarrow{r_{k-1}},~k=1,...,N.
\end{equation}
\begin{figure}
\includegraphics[width=13.5cm,clip]{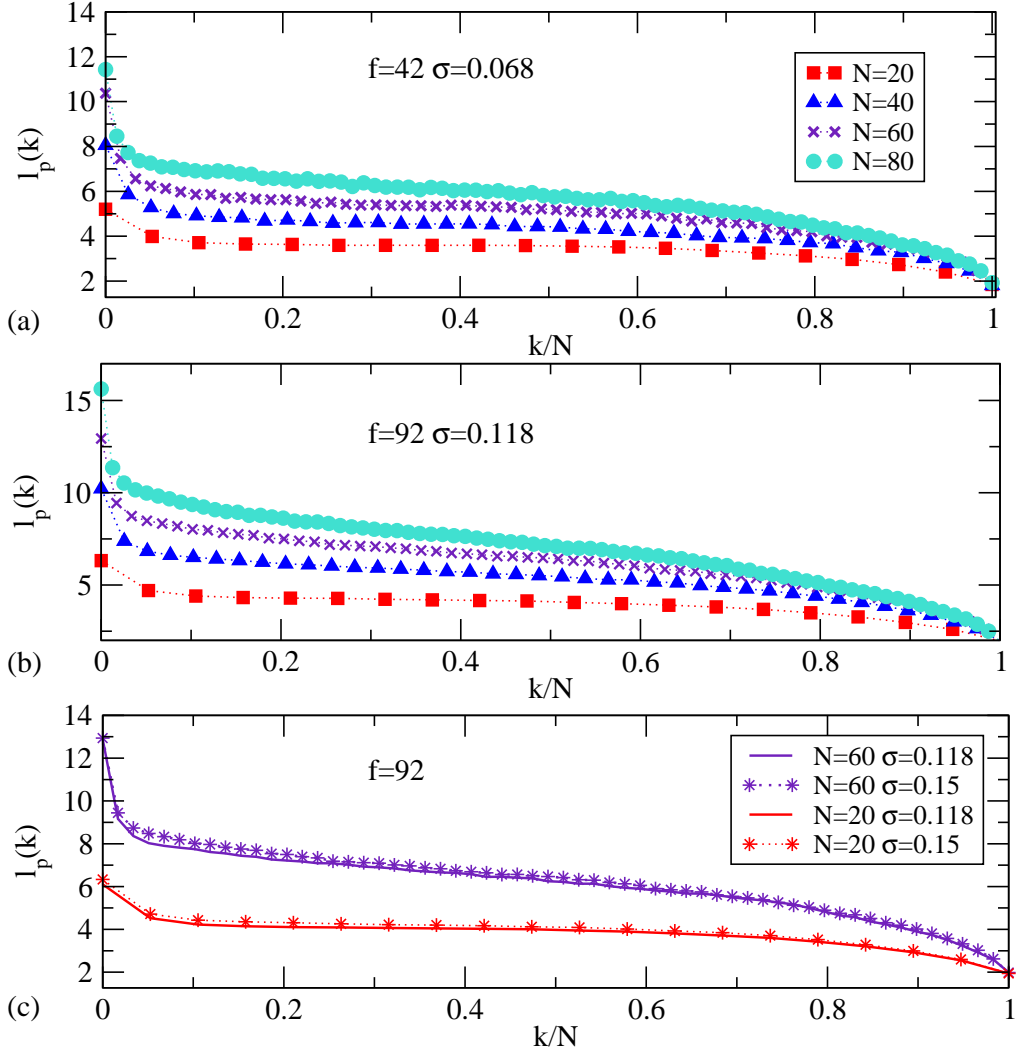}
\caption{\label{fig4}Local persistence length (as a measure of intrinsic chain orientation along the end-to-end vector) $l_p(k)$, Eq.[(6)], plotted vs. the normalized monomer index $k/N$ for the case $f$=42(a) and 92(b).
Case (c) compares for $f$=92 two grafting densities ($\sigma$=0.118 for $R_c$=7.9 and $\sigma$=0.15 for $R_c$=7, respectively) to demonstrate that the dependence of $l_p(k)$ on grafting density is very weak, at least in the range of parameters investigated in this paper.}
\end{figure}
Here $\overrightarrow{r_k}$ denotes the position of the $k$'th monomeric unit along a chain, starting with
$k$=0 at the grafting site. While for free chains (under good solvent conditions) $l_p(k)$ is known to exhibit a broad maximum near $k$=$N/2$, which scales as\cite{Schaefer_2004,Hsu_2010} $l_{p,max}(k)\propto N^{2\nu -1}$, we find here (Fig.\ref{fig4}) that $l_{p,max}(k)$ occurs near the grafting site, for $k$=1, and decreases monotonically with $k$, getting of order unity at the free chain end.
We recall here that
we work with a fully flexible model, lacking any intrinsic stiffness of "chemical origin".
Moreover under good solvent conditions we can not have quantitatively precise information on the persistence length as calculated from Eq.~\ref{pl} (see Ref.~\citenum{Hsu_2010}). 
However, Eq.~\ref{pl} still gives a meaningful information on the relative orientation of bond vectors, correlated with the end-to-end vector orientation. Nevertheless 
Fig.~\ref{fig4} demonstrates that the behavior is mostly controlled by the excluded volume interactions between the monomers 
of the single 
chain;
the repulsion due to the interchain interactions does not yet lead to a significant increase in orientation of the chains. 
On the other hand the chain linear size 
grows moderately in the radial
directions as the grafting density is increased.
\begin{figure}
\includegraphics[width=15.0cm,clip]{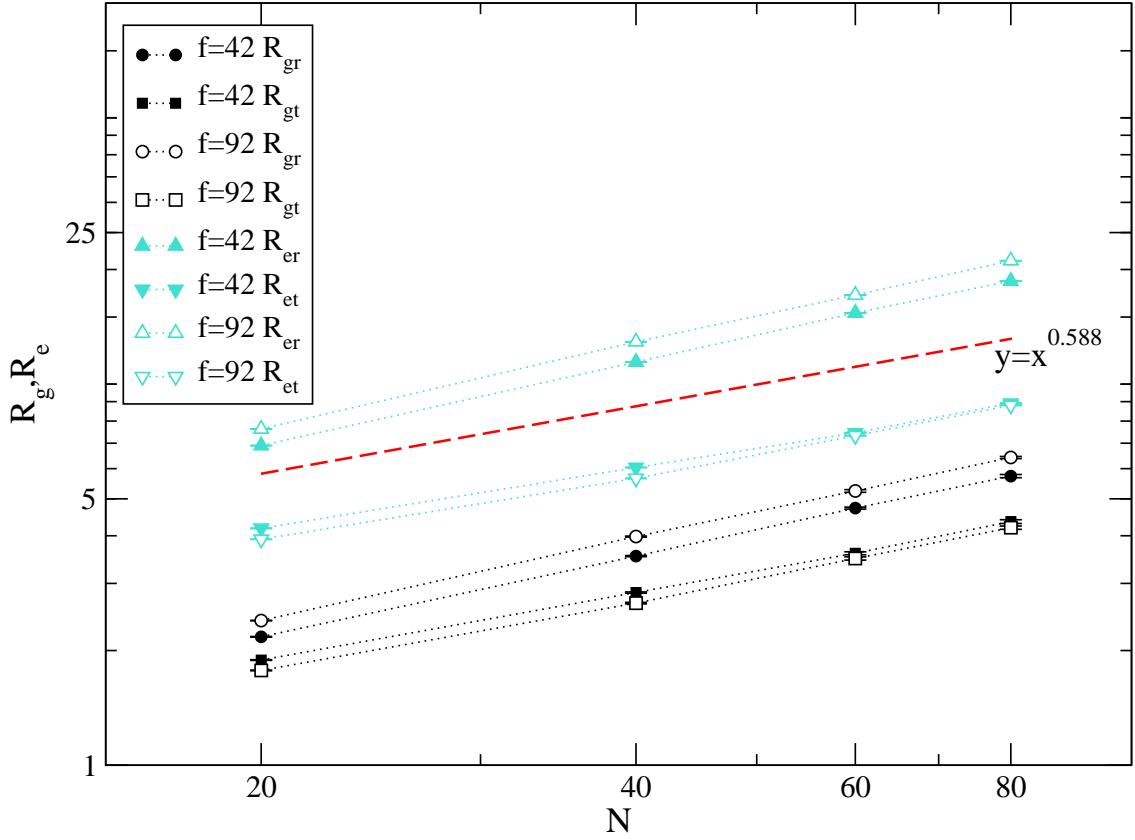}
\caption{\label{fig5}Log-log plot of mean square radius and end-to-end distance components in radial and tangential directions versus chain length.
Here $R_{gr}$=$\sqrt{\langle R_{gz}^2\rangle}$,  $R_{gt}$=$\sqrt{\langle R_{gx}^2 +R_{gy}^2\rangle/2 }$, $R_{er}$=$\sqrt{\langle R_{gz}^2\rangle}$, $R_{et}$=$\sqrt{\langle R_{ex}^2 +R_{ey}^2\rangle/2}$ orienting the $z$ axis from the center of mass of a chain perpendicular to the grafting surface of the sphere, and the $x,y$ axes are oriented perpendicular to this $z$-axis in each chain configuration that is analyzed.
Two choices of $f$ are shown, $f$=42 and
92.}
\end{figure}
In Fig.~\ref{fig5} we show as an example the log-log plot of mean square radius and end-to-end distance components in radial and tangential directions versus $N$, for $\sigma$=0.068 and $\sigma=0.118$.
The tangential components decrease with increasing $f$, 
in agreement with the increase of the stretching of the chains with the grafting density.
Of course the differences reduce as the number of monomers grows.
Note, however, that all the chain linear dimensions still scale with chain length according to the standard excluded volume power law, $R$$\propto$$N^\nu$. Neither the scaling relation for star polymers \cite{Daoud_1982}
%
%
\begin{equation}\label{RGr}
\langle R_{gr}^2\rangle \propto \langle R_{er}^2\rangle \propto N^{2\nu} f^{1-\nu}
\end{equation}
nor for chains in planar brushes in the strong stretching limit\cite{Milner_1988} 
\begin{equation}\label{RGz}
\langle R_{gz}^2\rangle \propto \langle R_{ez}^2\rangle \propto N^{2\nu} \sigma^{2/3}
\end{equation}
describe our numerical results 
perfectly.
Thus, the conclusion again is that there exists a very extended regime of gradual crossover from the planar brushes to the star polymer limit.
The results we obtained by MD simulations probe only a small section of this regime, as noted above.

It is worth pointing out that $R_{gr}$ and $R_{gt}$ are of the same order of magnitude; if one would take blob pictures like that of Daoud and Cotton\cite{Daoud_1982}
literally, one would conclude  each chain ``lives" in an angular sector of extent $4\pi/f$, and hence for large $f$ one would have $R_{gt}<<R_{gr}$.
Obviously, for entropic reasons it is more favorable that the different chains share these angular sectors, and thus the chains are also less stretched in radial directions than expected from the Daoud Cotton picture.

\begin{figure}
\includegraphics[width=14.5cm,clip]{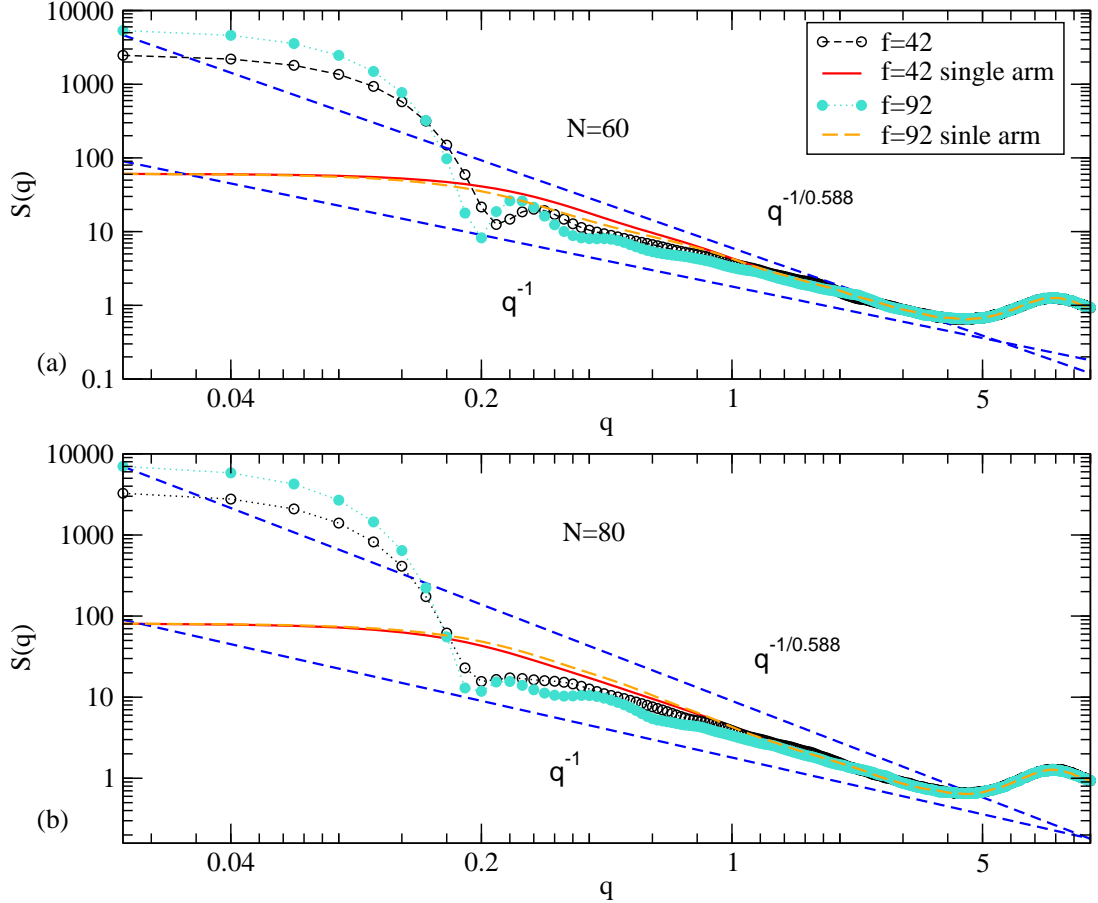}
\caption{\label{fig6}Log-log plot of the structure factor of a spherical brush (empty circles and full circles) and of a single arms (full curves without symbols) plotted versus wavenumber $q$, in the case $R_c$=7, $\sigma$=0.068, $f$=42, and $R_c$=7.9, $\sigma$=0.118, $f$=92, for two
chain lengths: $N$=$60$(a) and $N$=80(b).
Broken straight lines indicate the power law $S(q)\propto q^{-1}$ or $S(q)\propto q^{-1/\nu}=q^{-1/0.588}$, respectively.}
\end{figure}

As a last point of this section, we consider the structure factor describing the total scattering from the spherical brush, as well as the structure factor describing the spherically averaged scattering intensity from single arms (Fig.\ref{fig6}).
While the latter resembles the structure factor of an isolated polymer under good solvent conditions {crossing over from $S(q)=N(1-q^2<R_g^2>/3)$ at small $q$ to $S(q)\propto q^{-1/\nu}$ at larger $q$ \cite{deGennes_1979}}, the structure factor for the full spherical brush clearly exhibits a more interesting behavior. It resembles the scattering from a compact sphere at small $q$ and exhibits a minimum, the location of which is given roughly by $1/R_c$; the maximum at larger $q$ is more and more washed out as $N$ increases. These data clearly resemble quantitatively experimental scattering data from spherical micelles.\cite{McConnell_1994,Foenster_1996}

\section{Comparison to Results from Density Functional Calculations and to Other Related Works}\label{section4}
In this section we mainly focus on three different aspects:
the limit of applicability of the MD method, the density functional theory as a possible approach to extend the accessible  range of parameters,
the peculiarities of spherical brushes (crossover regime) with respect to star polymers and planar brushes.

In view of the rather large computational efforts  needed in order to generate well equilibrated MD data on spherical polymer brushes, with very good statistical accuracy and varying a number of parameters, it is worthwhile to ask whether or not one can obtain results of comparable quality using a ``cheaper" semi-analytical method, such as density functional theory (DFT).\cite{Egorov_2005,*Patel_2005,*Egorov_2007,*Egorov_2007_b,*Striolo_2007,Woodward_1990,*Woodward_1990_b} 
\begin{figure}
\includegraphics[width=13.0cm,clip]{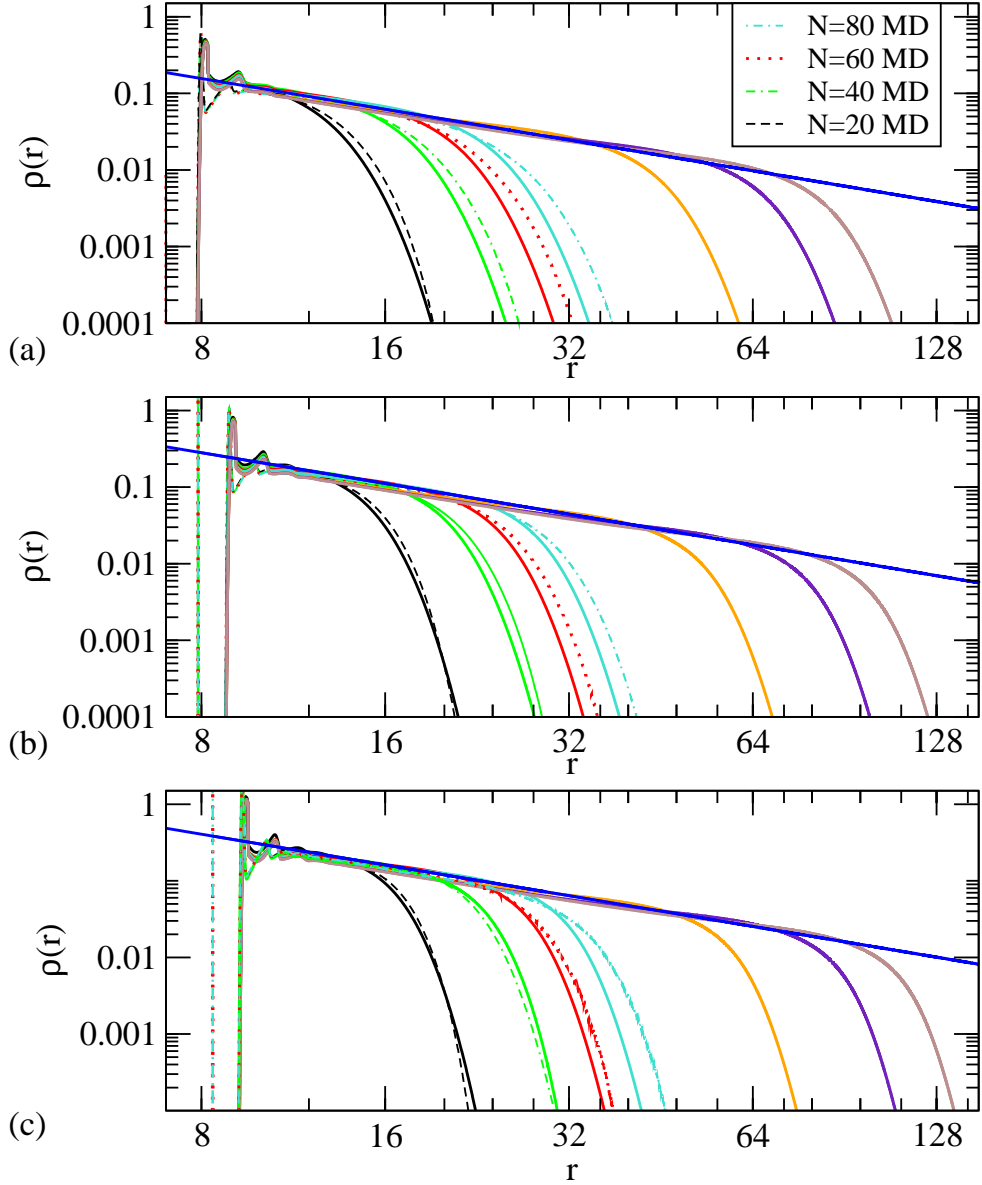}
\caption{\label{fig7}Log-log plot of the monomer density $\rho(r)$ versus $r$ for $f$=42(a) 92(b), 162(c) comparing MD results (broken lines) with DFT results (full lines), for $N$=20, 40, 60 and 80 (from left to right). In the picture the results for $N$=250, 500, 750 obtained via DFT are also shown (the three right most curves).}
\end{figure}
\begin{figure}
\includegraphics[width=14.0cm,clip]{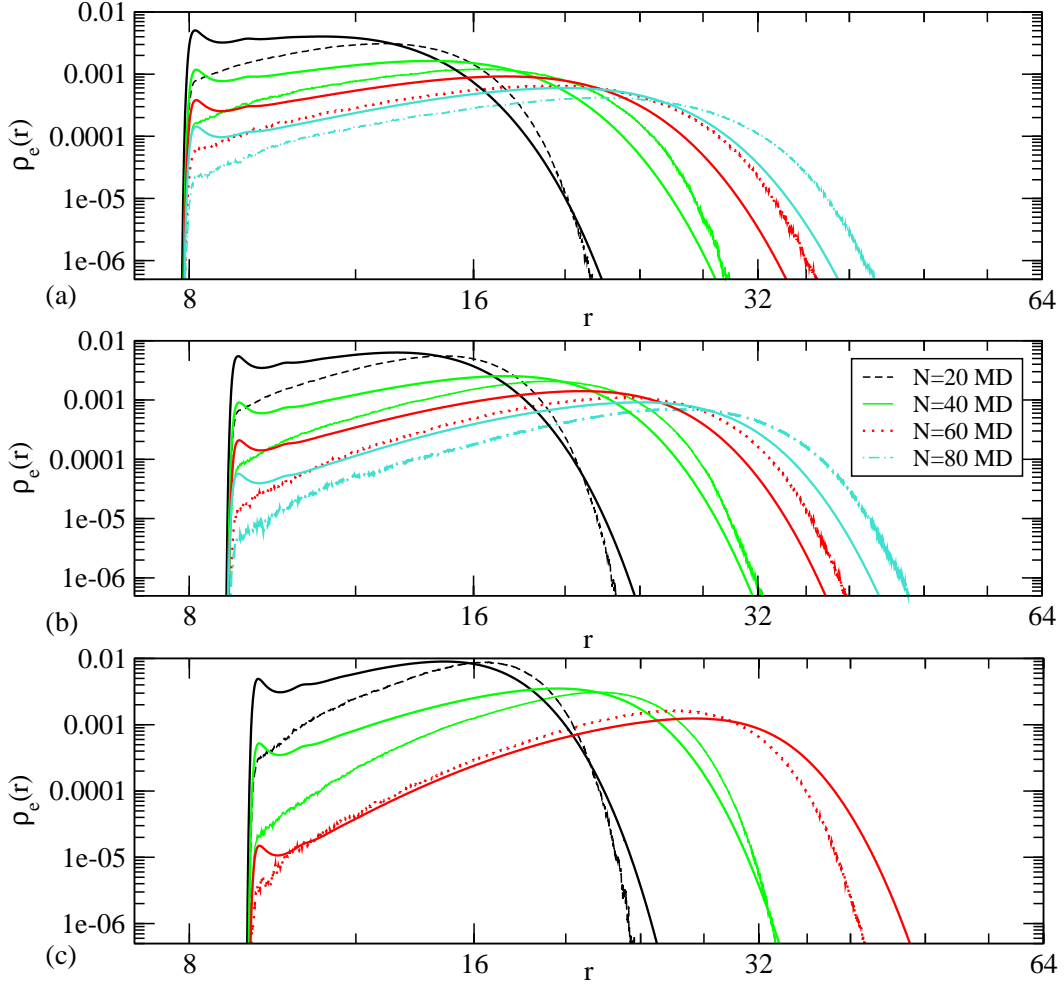}
\caption{\label{fig8}Log-log plot of the free end monomer density $\rho_e(r)$ versus $r$ for $f$=42(a) 92(b), 162(c) comparing MD results (broken lines) with DFT results (full lines), for $N$=20, 40, 60 and 80 (from left to right). }
\end{figure}

Since the details of these calculations have been given in a different context elsewhere \cite{Egorov_2005,*Patel_2005,*Egorov_2007,*Egorov_2007_b,*Striolo_2007} we here describe the results only, but we emphasize that a perfect agreement can not be expected, since the DFT uses a 
slightly different model for the chain molecules, employing tangent hard spheres rather than FENE+LJ potential (Eqs.~\ref{LJ},\ref{FENE})
to create chain connectivity. As a result, the chains molecules modelled by DFT are slightly different that those studied by MD, leading to a slight mismatch between the results of both methods.
Nevertheless one recognizes that the DFT reproduces the general trend of the MD results rather faithfully (Fig.~\ref{fig7},~\ref{fig8}). Also the layering of $\rho(r)$ close to the grafting surface is accurately reproduced (Fig.~\ref{fig7}).

\begin{figure}
\includegraphics[width=14.0cm,clip]{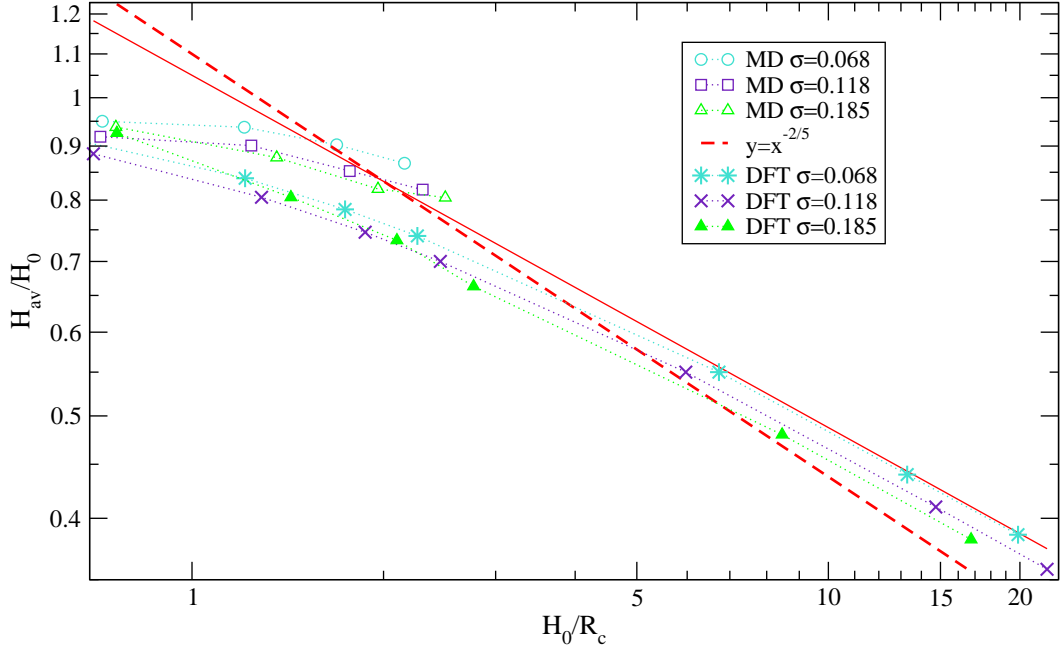}
\caption{\label{fig9} Log-log plot of the normalized brush height $H_{av}/H_0$
versus the normalized inverse radius $H_0/R_c$, comparing MD results (open symbols) to DFT results (filled triangles, $\sigma=0.185$; crosses, $\sigma=0.118$; asterisks, $\sigma=0.068$).
The theoretical asymptotic slope predicted from SCFT (-2/5) is shown as a broken straight line.
The DFT results converge to a power law with slope of minus one third (or -2/6), 
(full straight line).
Note that $H_{av}$ was computed from Eq.~\ref{Hav} while the corresponding height $H_0$ of a planar brush, using the same values of chain length $N$ and grafting density $\sigma$, we calculated from Eq.~\ref{H0}.}
\end{figure}

In order to study the crossover of the average brush height $H_{av}$ from the flat brush limit to the star polymer limit in detail, we follow Wijmans and Zhulina \cite{Wijmans_1993} by defining $H_{av}$ from the second moment of the monomer density profile as follows
\begin{equation}\label{Hav}
H_{av}^2=\int_{R_c}^\infty r^2\,dr \rho(r)(r-R_c)^2/\int_{R_c}^\infty r^2\,dr \rho(r)
\end{equation}
The limiting height for a planar brush, obtained from Eq.~\ref{Hav} as $R_c\rightarrow\infty$, is denoted as
\begin{equation}\label{H0}
H_{0}^2=\int_{0}^\infty \rho(z)z^2\,dz/\int_{0}^\infty dz\,\rho(z)
\end{equation}
From the SCFT treatment of Wijmans and Zhulina \cite{Wijmans_1993}
we conclude that in the limit $R_c\rightarrow\infty$, $N\rightarrow\infty$ (these limits need to be taken together such that $0<H_0/R_c<\infty$) a scaling property holds, where $f(\frac{H_0}{R_c})$ is some scaling function,
\begin{equation}\label{Hav/H_0}
H_{av}/H_0=f(H_0/R_c)
\end{equation}
irrespective of $R_c$ and $N$.
The scaling is also irrespective of the grafting density $\sigma$, provided it is chosen such that the monomer density in the brush stays in the semidilute regime:
the crossovers to either the mushroom regime or to dense brushes where $\rho(r)$ attains melt-like densities near the grafting surface are not covered.

As it is well-known, SCFT in the flat brush regime predicts that \cite{Milner_1991,Halperin_1991,Grest_1995,Szleifer_1996,Grest_1999,Leger_1999,PolymerBrushes_2004}
\begin{equation}\label{H0_brush}
H_{0}=\sigma^{1/3}N
\end{equation}
On the other hand, the star polymer scaling limit \cite{Daoud_1982}
implies that $H_{av}\propto N^{3/5}$.
As a consequence, the scaling function $f(X)$
 for large $X$ must behave as $f(X)\propto X^{-2/5}$, and hence one predicts \cite{Birshtein_1984,Birshtein_1986,Wijmans_1993,Toral_1993}
%
\begin{equation}\label{eq13}
H_{av}\propto H_{0}^{3/5} R_c^{2/5}\propto\sigma^{1/5}N^{3/5}R_c^{2/5}.
\end{equation}
Fig.\ref{fig9} presents a plot of our MD data, using such a scaling representation, and compares them to the corresponding DFT results.
The latter method has the distinctive advantage that it can be applied to much longer chains.
%
%
%
%
\begin{figure}
\includegraphics[width=13.0cm,clip]{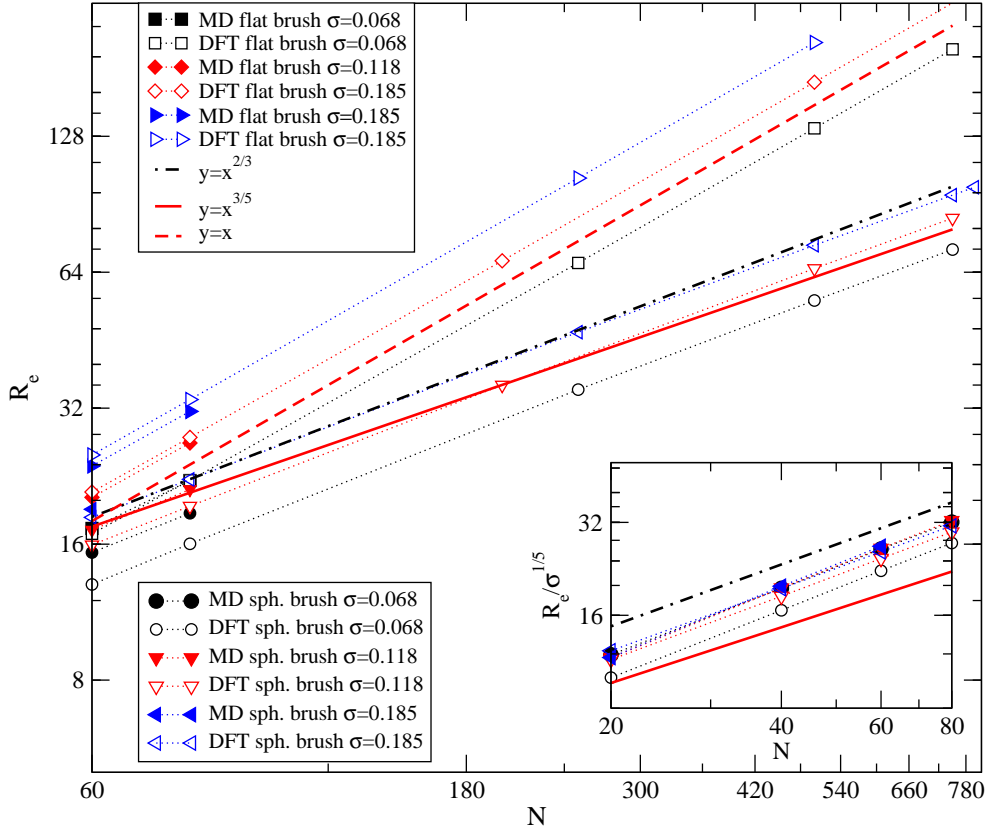}
\caption{\label{fig10} Log-log plot of the end-to-end distance of polymers grafted in flat and spherical brushes as obtained from MD (full symbols) and DFT (open symbols);
straight lines indicate exponents $1$ (broken straight line), $2/3$ (dash-dotted straight line) and $3/5$ (full line), respectively.
Inset: Log-log plot of the end-to-end rescaled distance $R_e/\sigma^{1/5}$ versus the number of monomers per chain, for $N<$80.}
\end{figure}
While (for our choices of $R_c$) the MD data do not extend beyond $H_0/R_c\approx 2.5$, the DFT data can easily be taken to $H_0/R_c\approx 20$, using chain lengths up to $N$=750, which would be inaccessible for our MD study.
As expected from the comparison of the profiles, DFT and MD results are close to each other, and exhibit a similar trend.
However, it is also clear that our data do not fall in a regime where the scaling described by Eq.~\ref{Hav/H_0} is valid:
the results for the three grafting densities used in our MD work do not superimpose on a single "master curve",which then would yield an estimate of a piece of the scaling function $f(X)$.
It is also clear that the MD data did not reach the regime where Eq.~\ref{eq13}
is valid.
The DFT results, on the other hand, do reach an asymptotic power law but it is inconsistent with  Eq.~\ref{eq13},
rather implying that $f(X)\propto X^{-1/3}$.
We lack an explanation as to why DFT predicts a star polymer scaling $R_{star} \propto N^{2/3}$ (following from $f(X)\propto X^{-1/3}$) rather than $R_{star}\propto N^{3/5}$.

These discrepancies are also apparent when we study the end-to-end distance of the chains, predicted by DFT for both flat and spherical brushes (Fig.~\ref{fig10}):
while for flat brushes DFT predicts correctly $R_e\propto N$,
for spherical brushes one finds $R_e\propto N^{2/3}$ rather than the scaling result, $R_e\propto N^{3/5}$.
Thus DFT is less useful for spherical brushes than for flat brushes, overestimating 
the chain stretching in spherical brushes for large $N$ systematically.

\begin{figure}
\includegraphics[width=12.0cm,clip]{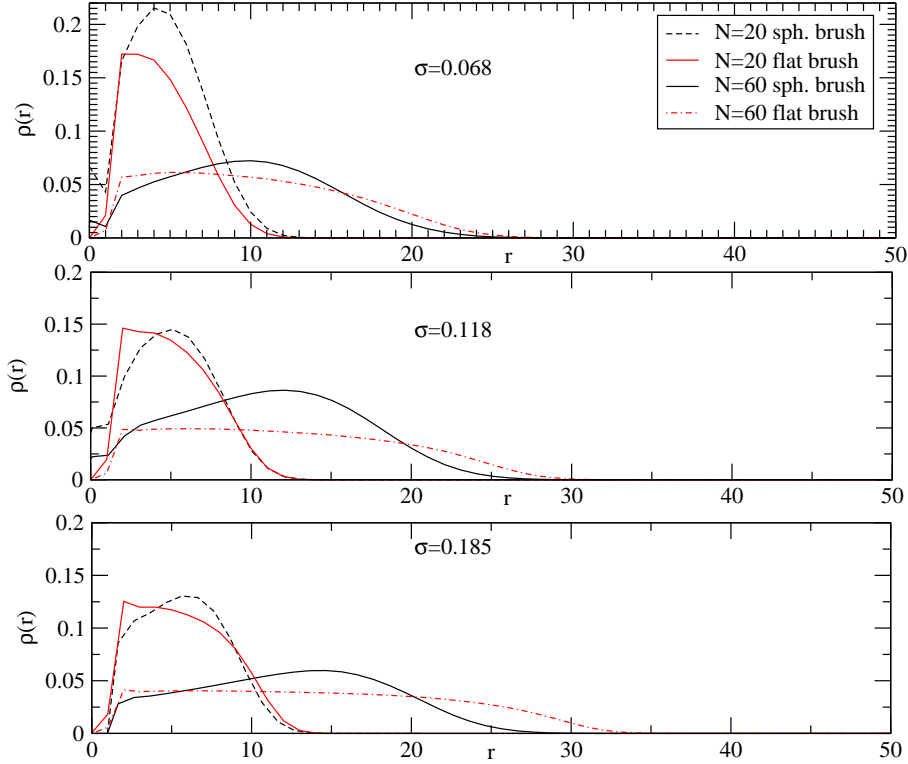}
\caption{\label{fig11} Monomer density profiles $\rho(r)$
for $N$=20 and $N$=60 plotted versus distance $r$ from the grafting surface for both spherical brushes and flat brushes, as indicated, including three grafting densities:
$\sigma=0.068$ (upper panel), $\sigma=0.118$ (middle panel), $\sigma=0.185$ (lower panel). All the profiles are normalized to unit area
$\int_{0}^\infty dz\,\rho(z)$=1 (flat brushes) or $\int_{R_c}^\infty dr\,r^2\,\rho(r)$=1 (spherical brushes), respectively.}
\end{figure}

It is also of interest to directly compare the density profiles as obtained from MD for flat brushes and for spherical brushes (Fig.~\ref{fig11}).
As expected, for $N=20$ the differences between the profiles for flat and spherical brushes are rather minor, since for the chain values of $R_c$ then $H_{av}<<R_c$ for all these grafting densities.
On the other hand, for $N=60$ the curvature of the grafting surface has already a clearly recognizable effect on the profile.

\begin{figure}
\includegraphics[width=12.0cm,clip]{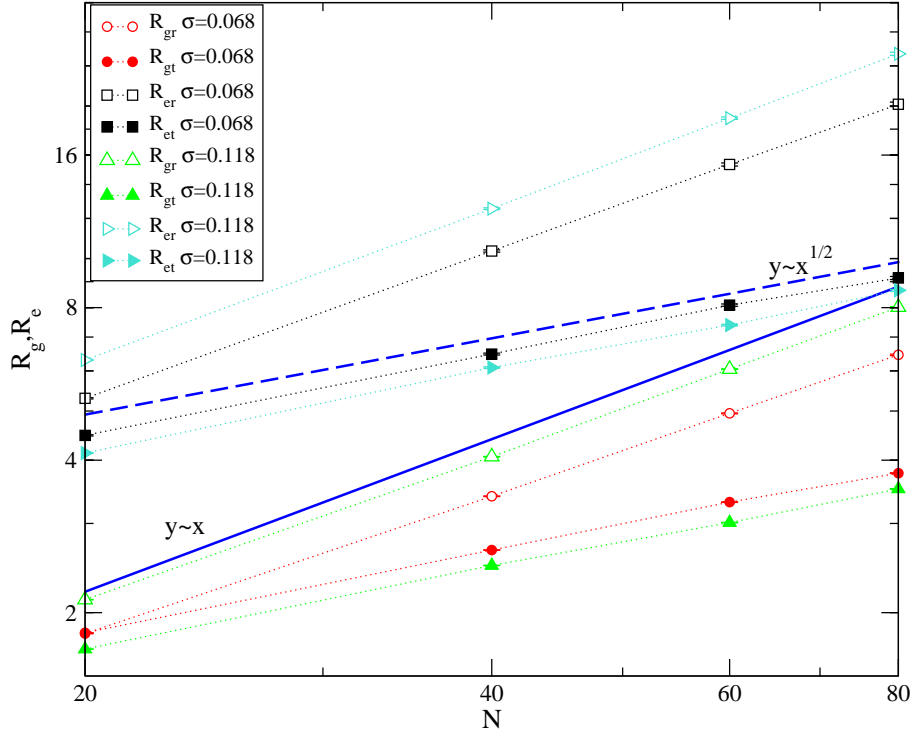}
\caption{\label{fig12}Log-log plot of mean square radius and end-to-end distance components in radial and tangential directions versus chain length.
Here $R_{gr}$=$\sqrt{\langle R_{gz}^2\rangle}$,  $R_{gt}$=$\sqrt{\langle R_{gx}^2 +R_{gy}^2\rangle/2 }$, $R_{er}$=$\sqrt{\langle R_{gz}^2\rangle}$, $R_{et}$=$\sqrt{\langle R_{ex}^2 +R_{ey}^2\rangle/2}$ orienting the $z$ axis perpendicular to the flat grafting surface, and the $x,y$ axes lie on the plan parallel
to the surface in each chain configuration that is analyzed.
Two choices of $f$ are shown, $f$=42 and
92.}
\end{figure}

For completeness we plot 
the radial and tangential components of the radius of gyration and end-to-end distance of the single polymer grafted to the flat brush.
The figure shows the different scaling trends with respect to the spherical brush cases (see Fig.~\ref{fig12} for comparison).

Finally we mention that in the recent experiment of Dukes $et~ al.$\cite{Dukes_2010} a scaling $H_{av}\propto N^{4/5}$ was observed, eventually crossing over to $H_{av}\propto N^{3/5}$ at larger $N$.
The authors emphasize, however, that this behavior should not be explained in terms of Eq.~\ref{Hav/H_0} since their brushes are in a regime where near the grafting surface the semidilute concentration regime is distinctly exceeded,
 and one must build the description on a theory for concentrated rather than semidilute polymer brushes.
Nevertheless they find that the approximate prediction of 
Wijmans and Zhulina\cite{Wijmans_1993}
for the crossover grafting function $f(X)$ roughly accounts for their data, 
as well as for previous data of other authors \cite{Savin_2002,Ohno_2007}.
However, it is clear that a fully satisfactory theoretical description of these data does not yet exist.
Nevertheless it is encouraging that experiments have become feasible for conditions that are not so different from the conditions studied in the present work.

\section{Conclusions}\label{section5}
In this work we have presented a detailed and quantitative analysis of conformation and structural properties of spherical brushes.

In the first instance 
we focused  on molecules where 
the core has a radius comparable to the polymer linear dimension.
In particular we have carried out monomer resolved Molecular Dynamic simulations
in order to study the monomer density profile and the distribution of the free end monomers
for three values of the grafting density (namely $\sigma$=0.068,0.118,0.185).
In order to clarify the origin of the
inhomogeneous stretching of the chains, and the development of multiple 
relevant
 length scales 
we calculated the local persistence length, the structure factor of single chain and full molecule
 as well as the radial and tangential components of the radius of gyration and end-to-end distance of the single polymer.
In the regime of parameters investigated the structure of the brush is mostly driven by the intrachain excluded volume
interactions.
The chains are less stretched in radial directions than expected from the Daoud Cotton picture.
The main conclusion here is that there exists an extended regime of gradual crossover between star polymer and 
flat brush behavior.
Using the ratio of the brush height at a flat surface to the core radius
as a variable for this crossover, we can state that this crossover extends over several decades in this variable, in order to cover both star polymer scaling and flat brush scaling relations.
In addition, one has to reach the strong stretching limit of semidilute polymer brushes (avoiding both mushroom-like behavior and too dense melt-like brushes) to observe this simple universal crossover scaling behavior.
Both by simulations and in typical experimental systems it is not easy to reach this asymptotic scaling regime, and hence the crossover scaling considerations are only a rough guidance to interpret the expected behavior.

We also comment on the use of the present model to interpret scattering from spherical micelles.
Indeed the chain lengths and total number of chains grafted to the spherical core particle are in a similar range as for typical micelles in experimental studies. While the scattering curves found here (\ref{fig6}) are quantitatively similar to experimental data, we do expect some differences due to the fact that in micelles there is a smooth interfacial profile between the monomers of type A in the collapsed core and the monomers of type B in the (swollen) corona
(referring here to block copolymer micelles of composition 
$A_{l}B_{1-l}$ with $l<<$1 
in a selective solvent, poor for A but good for B, as used in typical experiments). Therefore one cannot expect the pronounced layering in the density profile of the corona monomers, that we see here for radii $r$, near the radius $R_c$ of the hard particle to which our chains are grafted (\ref{fig-1}).

In the second instance we  considered the density functional approach 
as a possible alternative method.
The results obtained via DFT are in a very good agreement with our simulation results 
and open up the possibility to extend the range of parameters investigated.
Nevertheless the theory does not reproduce exactly the star polymer scaling.
Moreover comparing the results obtained on flat brushes via
MD simulations and DFT 
we observed that
 DFT is more precise in describing 
the scaling of flat than spherical brushes, overestimating 
the chain stretching as the radius of curvature of the surface increases.

We finally compare the MD data for flat and spherical brushes
in order to emphasize the peculiarity 
of the regime 
characterized by 
a grafting surface with a radius comparable to the polymer linear dimension.

This ¨crossover¨ regime of the surface curvature radius 
(between flat interfaces or star-like polymers)
is of great interest for many practical purpose closely related to our daily life.
One aspect we are interested in is related to the
investigation of the adsorption of proteins to polymeric spherical brushes, 
e.g. how adsorbed or chemically-attached polymers affect the interaction between nanoparticles and globular proteins.
A full understanding of the structural properties of the polymeric unit is a first 
necessary steps in a such direction.

%
%

%

\begin{acknowledgments}
One of use (S.E.) acknowledges partial support by the Alexander von Humboldt-Foundation via a research fellowship, and another (A.M.) acknowledges partial support from the Deutsche Forschungsgemeinschaft (DFG), grant $N^o$ Bi $314/22$ and SFB 625/A3.
Both S.E. and A.M. are grateful to the Institut f{\"u}r Physik of the Johannes Gutenberg Universit{\"a}t Mainz for the hospitality during their visit at Mainz.
We are also grateful to the J{\"u}lich super computer center for allocation of computer time at the SOFTCOMP and JUROPA computers.
F.LV thanks Hsiao-Ping Hsu and Christian Mayer for helpful discussions.
\end{acknowledgments}

\providecommand{\noopsort}[1]{}\providecommand{\singleletter}[1]{#1}%

\end{document}